\documentclass[12pt]{article}
\usepackage{amssymb}
\usepackage{amsmath}
\usepackage{epsfig}
\usepackage{float}
\newcommand{\be}{\begin{equation}}
\newcommand{\ee}{\end{equation}}
\newcommand{\bea}{\begin{eqnarray}}
\newcommand{\eea}{\end{eqnarray}}
\begin{document}

\begin{center}
{\bf NEUTRINO MASSES, MIXING AND OSCILLATIONS}\footnote {A report at the International School of 
Nuclear Physics ``Neutrino in Cosmology, in Astro, Particle and Nuclear Physics'' Erice, Italy, Sept. 16-24, 2005}

\end{center}

\begin{center}
S. M. Bilenky
\end{center}
\begin{center}
{\em  Joint Institute
for Nuclear Research, Dubna, R-141980, Russia, and\\
Scuola Internazionale Superiore di Studi Avanzati,
I-34014 Trieste, Italy.}
\end{center}

\begin{abstract}
Basics of neutrino oscillations is discussed. Importance of time-energy uncertainty relation is stressed. 
Neutrino oscillations in the leading approximation and evidence for neutrino 
oscillations are briefly summarized.
\end{abstract}

\section{Introduction}
Evidence for neutrino oscillations obtained in the Super-Kamiokande \cite{SK}, SNO \cite{SNO}, 
KamLAND \cite{Kamland} 
and other neutrino experiments \cite{K2K,Cl,Gallex,Sage,SKsol,Soudan,Macro}
 is one of the most important recent discovery in particle physics. There are no natural explanations of the smallness of neutrino masses in the Standard Model. 
A new, beyond the Standard Model mechanism of the generation of neutrino masses 
is necessary. In order to reveal the  origin of the discovered phenomenon 
new experimental data are definitely needed.

We will discuss here first the basics of the neutrino oscillations. Then we will consider neutrino oscillations in the leading approximation and give a brief summary of the data. In conclusion we will make some remarks about the possible future of the physics of massive and mixed neutrinos.

\section{Basics of neutrino oscillations}

Investigation of neutrino oscillations is based on the following  assumptions:

\begin{enumerate}
\item
Neutrino interaction is given by the standard CC and NC Lagrangians
\be
\mathcal{L}_{I}^{\mathrm{CC}}(x) = - \frac{g}{2\sqrt{2}} \,
j^{\mathrm{CC}}_{\alpha}(x) \, W^{\alpha}(x) + \mathrm{h.c.};~~
j^{\mathrm{CC}}_{\alpha}(x) =2\, \sum_{l=e,\mu, \tau} \bar \nu_{lL}(x)
\gamma_{\alpha}l_{L}(x)\label{1a}
\ee
and
\be
\mathcal{L}_{I}^{\mathrm{NC}}(x) = - \frac{g}{2\cos\theta_{W}} \,
j^{\mathrm{NC}}_{\alpha}(x) \, Z^{\alpha}(x);~~j^{\mathrm{NC}}_{\alpha}(x) =\sum_{l=e,\mu, \tau} \bar
\nu_{lL}(x)\gamma_{\alpha}\nu_{lL}(x).\label{1b}
\ee

Here  $g$ is the SU(2) interaction constant and $\theta_{W}$ is the weak angle.

\item  $\nu_{lL}(x)$ ($l=e,\mu, \tau $)  in (\ref{1a}) and (\ref{1b})  are ``mixed'' fields

\be
\nu_{lL}(x)=\sum_{i=1}^{3} U_{li}\,\nu_{iL}(x),\label{3}
\ee
where $\nu_{i}(x)$ is the field of neutrino with mass $m_{i}$ and $U$ is the unitary PMNS \cite{Pont,MNS}
mixing matrix.\footnote{We assumed that the number of massive neutrinos is equal to the number of flavor neutrinos. All existing neutrino oscillation data,
 with the exception of the  data of the LSND experiment\cite{LSND}, 
are in a perfect agreement with this assumption. The LSND data are going to be checked by the running 
MiniBooNE experiment \cite{Miniboone}} 

\end{enumerate}

There are two possibilities for neutrinos with definite masses $\nu_{i}$:
\begin{itemize}
\item If total lepton number $L=L_{e}+ L_{\mu}+L_{\tau}$ is conserved, {\em $\nu_{i}$ are Dirac particles}
($L(\nu_{i})=1,  L(\bar\nu_{i})=- 1$).
\item
If $L$ is violated, {\em $\nu_{i}$ are Majorana particles}. The field $\nu_{i}(x)$ satisfies in this case 
Majorana condition
\be
\nu^{c}_{i}(x) =\nu_{i}(x),\label{4}
\ee
where $\nu^{c}_{i}(x)=C\,\bar\nu^{T}_{i}(x)$ ($C$ is the matrix of the charge conjugation).
\end{itemize}
Neutrino oscillations is the most important implication of the neutrino mixing (\ref{3})
We will discuss now briefly the basics of this phenomenon (for other discussions see \cite{Carlo} and numerous references therein).

Neutrinos are produced in weak processes. Let us consider the production of a neutrino 
 in a CC weak decay
\be
a\to b +l^{+}+\nu_{i},\label{5}
\ee
where $a$ and $b$ are some hadrons. For final neutrino state we have
\be
 |\nu_f\rangle= \sum^{3}_{i=1} |\nu_i\rangle  \,~ \langle \nu_i \,l^{+}\,b\,| S
|\,a\rangle, \label{6}
\ee
where $ |\nu_i\rangle$ is the state of neutrino with mass $m_{i}$, momentum $\vec{p}$ and energy 
$E_{i}=\sqrt{p^{2}+m^{2}_{i}}$.

Because of the Heisenberg uncertainty relation production of different 
$\nu_{i}$ can not be revealed in the  processes in which usual neutrino beams with neutrino energies 
$\gtrsim$ MeV are produced.  
Thus, we have
\be
\langle \nu_i \,l^{+}\,b\,| S |\,a\rangle= U^{*}_{li}\,~\langle
\nu_l \,l^{+}\,b\,| S |\,a\rangle_{SM},\label{8}
\ee
where $\langle \nu_l \,l^{+}\,b\,| S |\,a\rangle_{SM}$ is the standard model matrix element of the process 
$a\to b +l^{+}+\nu_{l}$.\footnote{ Because neutrino masses  are much smaller than 
neutrino energies,  neutrino masses  can be neglected in 
the matrix elements $\langle \nu_l \,l^{+}\,b\,| S |\,a\rangle_{SM}$ and in the corresponding 
phase space factors.
In 
tritium neutrino mass experiments, in which high-energy part of the $\beta$-spectrum is studied,
energies of neutrinos are comparable with expected neutrino masses. It is obvious
that in corresponding expression for the $\beta$-spectrum 
effect of neutrino masses must be taken into account} 

From (\ref{6}) and (\ref{8}) for the normalized neutrino state we have
\be
|\nu_{l}\rangle =\sum^{3}_{i=1}U^{*}_{li}\,|\nu_{i}\rangle ~~l=e,\mu,\tau
\label{9}
\ee
Thus, the flavor neutrino states  
$|\nu_{l}\rangle $ are  
{\em coherent superposition} of states of neutrinos with definite masses.

If at $t=0$ flavor neutrino $\nu_{l}$ was produced, the neutrino state in vacuum at the time $t>0$ 
is  given by
\be
|\nu_{l}\rangle_{t}=
e^{-H_{0}\,t}\,|\nu_{l}\rangle=\sum_{i}
|\nu_{i}\rangle\,e^{-iE_{i}t}\,U^{*}_{li},
\label{10}
\ee
where $H_{0}$ is the free Hamiltonian.

Neutrinos are detected via the observation of weak processes. Let us consider the inclusive process
\be
\nu_{i}+N \to l' + X.\label{11}
\ee
Neglecting small neutrino masses,  we have
\be
\langle l'\,X\,| S |\,\nu_{i}\,N \rangle=U_{l'i}\, \langle l'\,X\,|
S |\,\nu_{l'}\,N \rangle_{SM},\label{12}
\ee
where $\langle l'\,X\,|
S |\,\nu_{l'}\,N \rangle_{SM} $ is the SM matrix element of the process 
$\nu_{l'}+N \to l' + X$. From (\ref{10}) and (\ref{12}) for the normalized probability of 
$\nu_{l} \to \nu_{l'}$ transition in vacuum we obtain the following expression
\be
{\mathrm P}(\nu_{l} \to \nu_{l'}) =|\,\sum_{i}U_{l'  i} \,~
e^{- i\,E_{i}t} \,~U_{l i}^*\, |^2 = \sum_{i,k}U_{l'
i}\,U^{*}_{l'  k} \,~ e^{- i\,(E_{i}-E_{k})\, t} \,~U_{l
i}^*U_{l'  k}\label{13}
\ee
This expression describes periodical transitions between flavor neutrino states with oscillation 
times given by 
\be
t_{ik}=\frac{2\pi}{|\Delta E_{ik}|};~
\Delta E_{ik}=  E_{i}-E_{k}=\frac{\Delta m^{2}_{ik}}{2p};~~\Delta m^{2}_{ik}= m^{2}_{k}-m^{2}_{i}.
~(i\not=k)
\label{14}
\ee
\section{Neutrino oscillations and time-energy uncertainty relation}
Let us consider
translations
\be x'=x+a, \label{16}
\ee
where $a$ is a constant vector.
In the case of the invariance under translations for  vectors of state and field operators we have
\be
|\Psi\rangle'=e^{iPa}\,|\Psi\rangle;~~
e^{-iPa}~O(x+a)~ e^{iPa}=O(x),
 \label{18}
\ee
where $P$ is the operator of the total momentum and   vectors $|\Psi\rangle$ and $|\Psi\rangle'$ 
describe the same physical state.

Let us apply now operator of the translation  $e^{iPa}$ to the flavor neutrino state $\nu_{l}$. 
We have
\be
|\nu_{l}\rangle'=e^{iPa}\,|\nu_{l}\rangle= e^{-i\vec{p}\vec{a}}
\sum_{l'} |\nu_{l'}\rangle\,\sum_{i} U_{l'i}\,e^{iE_{i}a^{0}}\,U^{*}_{li}.\label{19}
\ee
Thus, vectors $|\nu_{l}\rangle'$ and $|\nu_{l}\rangle$ describe  {\em different states}. 
This means that in the case of mixed states there is no invariance under translations and in transitions
 between different flavor states energy is not conserved.

This is connected with the fact that the states of the flavor neutrinos are non stationary states 
(see (\ref{10})). For such states time-energy uncertainty relation 
\be
\Delta E\,\Delta t \gtrsim 1\label{20}
\ee
takes place. In Eq. (\ref{20})
$\Delta t$ characterizes the rate of  the significant changes in the
system. In the case of the neutrino oscillations it is given by oscillation time (oscillation length).
\section{Neutrino oscillations in the leading approximation}
In the case of the three-neutrino mixing
$\nu_{l} \to \nu_{l'}$ transition probability  can be presented in the form
\be
{\mathrm P}(\nu_{l} \to \nu_{l'})  =|\delta_{l'l}+
\sum_{i=2,3} U_{l' i} \,~( e^{- i\,\Delta m^2_{1i} \frac {L}{2E}}-1)
\,~U_{l i}^*\, |^2,\label{21}
\ee
where $L\simeq t$ is the distance between neutrino production and detection points.

In the general case the transition probability 
${\mathrm P}(\nu_{l} \to \nu_{l'})$ 
is characterized by six parameters:
two mass-squared differences $\Delta m^2_{12}$ and $\Delta m^2_{23}$, three mixing angles
$\theta_{12}$, $\theta_{23}$ and $\theta_{13}$ and  one $CP$ phase $\delta$.
However, in the leading approximation
the picture of neutrino oscillations is greatly simplified (see \cite{BGG}).
This approximation is based on two 
inequalities which were found from analysis of experimental data
\begin{itemize}
\item $\Delta
m^2_{12} \simeq 3.3\cdot 10^{-2}~ \Delta m^2_{23}$ (all neutrino oscillation data)

\item $\sin^{2}\theta_{13}\ll 1$ (CHOOZ data \cite{Chooz})

\end{itemize}
If we neglect in the transition probabilities small terms proportional to 
$\frac{\Delta m^2_{12}}
{\Delta m^2_{23}} $ and $\sin^{2}\theta_{13} $
then from (\ref{21}) it follows that for $\frac{L}{E}\gtrsim \frac{1}{\Delta m^2_{23}}$
(atmospheric and accelerator long baseline experiments) dominant transitions are
$\nu_{\mu} \to \nu_{\tau}$ and 
$\bar\nu_{\mu}\to \bar\nu_{\tau}$. For the probability of 
$\nu_{\mu}$ ($\bar\nu_{\mu}$) to survive we obtain the standard two-neutrino expression
\be
{\mathrm P}(\nu_\mu \to \nu_\mu) ={\mathrm P}(\bar\nu_\mu \to \bar\nu_\mu)= 1 - \frac {1}
{2}\,\sin^{2}2\theta_{23}\, (1-\cos \Delta m_{23}^{2}\, \frac {L}
{2E}).\label{22}
\ee
In the solar and KamLAND reactor experiments, sensitive to small $\Delta m_{12}^{2}$, effect of the 
``large''
$\Delta m_{23}^{2}$ is averaged. For $\nu_{e}$ ($\bar\nu_{e}$) survival probability in vacuum 
(in matter) we obtain the following general expression \cite{Schramm}:
\be
{\mathrm P}(\nu_e \to \nu_e) ={\mathrm P}(\bar\nu_e \to \bar\nu_e)=
|U_{e3}|^{4}+(1-|U_{e3}|^{2})^{2}\,{\mathrm
P}^{(12)}(\nu_e \to \nu_e),\label{23}
\ee
where $|U_{e3}|=\sin \theta_{13}$ and 
${\mathrm P}^{(12)}(\nu_e \to \nu_e)$ is the two-neutrino $\nu_e$ ($\bar\nu_e$)  survival
probability in vacuum (in matter)  which depends on 
the oscillation parameters $\Delta m_{12}^{2}$ and $\sin^{2}\,\theta_{12}$.

If we neglect  $\sin^{2}\theta_{13}$,  for $\bar \nu_e$ survival probability in vacuum 
(KamLAND experiment) we obtain two-neutrino expression
\be
{\mathrm P}(\bar \nu_e \to \bar\nu_e)
=1-\frac{1}{2}~\sin^{2}2\,\theta_{12}~ (1 - \cos \Delta m_{12}^{2}
\,\frac {L}{2E})\label{24}
\ee
Let us notice 
that the second term in (\ref{24}) is the sum of (approximately equal) probabilities of 
the transitions $\bar \nu_e\to \bar \nu_{\mu}$   and 
$\bar \nu_e\to \bar \nu_{\tau}$. 

Thus, in the leading approximation three-neutrino oscillations are decoupled:
in the atmospheric-long baseline  and solar-KamLAND experiments
oscillations are described by two-neutrino type expressions which
depend, respectively, on  $\Delta m_{23}^{2}$, $\sin^{2}2\theta_{23}$ and $\Delta
m_{12}^{2}$,  $\tan^{2}\theta_{12}$.
Existing
experimental data are in  a good agreement with the leading approximation.

\section{Briefly on experimental data}
In the atmospheric Super-Kamiokande experiment \cite{SK} significant zenith angle asymmetry of the muon events was observed. For the integral up-down asymmetry of the muon events it was obtained
\be
\left(\frac{U}{D}\right)_{\mu}= 0.551\pm 0.035 \pm 0.004,\label{25}
\ee
where $U$ is the the total number  of the up-going muons (neutrino distances 13000 -500 km) and
$D$ is the the total number of the down-going muons
 (neutrino distances 10 - 500 km).
Recently clear oscillatory behavior
of  the $\nu_{\mu}$ survival probability as function of  $\frac{L}{L}$.
was demonstrated by the  Super-Kamiokande collaboration \cite{SK}.

From the two-neutrino analysis of all Super-Kamiokande data for the oscillation parameters the following 
90\% CL ranges were obtained
\be
1.5\cdot 10^{-3}\leq \Delta m^{2}_{23} \leq 3.4\cdot
10^{-3}\rm{eV}^{2};~~\sin^{2}2 \theta_{23}> 0.92.\label{26}
\ee
The Super-Kamiokande evidence  for neutrino oscillations was confirmed by the
accelerator K2K experiment \cite{K2K}. In this experiment
$\nu_{\mu}$  from KEK accelerator were detected in the Super-Kamiokande detector at
the distance 250 $\rm{km}$.
The expected number of $\nu_{\mu}$ events in the K2K experiment was equal to $151^{+12}_{-10}$.
107 $\nu_{\mu}$ events were observed.
The best-fit values of the oscillation parameters
obtained from analysis of the K2K data
\be
\Delta m^{2}_{23} = 2.8\cdot 10^{-3} \rm{eV}^{2};~~\sin^{2}
2\theta_{23}=1. \label{27}
\ee
are in agreement with (\ref{26}).

In the solar SNO experiment \cite{SNO} strong model independent evidence of $\nu_{e}$ disappearance was obtained. Solar neutrinos are detected in this experiment via the observation of the
reactions:
\be  \nu_e + d \to e^{-}+ p +p;~\nu_x+ d \to \nu_x + n +p;~\nu_x  + e \to \nu_x + e. \label{28}
\ee
Only high energy 
$^8 B$ neutrinos are detected in the SNO
experiment.
From the observation of the solar neutrinos through the detection of the CC
and NC reactions for the total fluxes of $\nu_{e}$ and 
$\nu_{e}, \nu_{\mu}, \nu_{\tau}$ it was found, respectively
\be
 \Phi_{\nu_{e}}^{\rm{SNO}} = (1.68 \pm 0.06 \pm 0.09)
\cdot 10^{6}\,~ cm^{-2}s^{-1}\label{29}
\ee
and 
\be 
\Phi_{\nu_{e,\mu,\tau}}^{\rm{SNO}} =(4.94 \pm 0.21 \pm 0.38
) \cdot 10^{6}\,~ cm^{-2}s^{-1}\label{30}
\ee
Thus, the total flux of $\nu_{e}$, $\nu_{\mu}$ and $\nu_{\tau}$ is
about three times larger than the flux of $\nu_{e}$.

In the reactor KamLAND experiment \cite{Kamland}
$\bar \nu_{e}$  from  53 reactors at  the 
average distance about 180 km from the detector  are detected via
the observation of the reaction $\bar \nu_{e}+p \to e^{+}+n$.
The expected number of the $\bar\nu_{e}$ events is equal to 
 $365.2 \pm
23.7$. 258 events were observed in the experiment.
For the ratio of the observed and expected events was found
\be
R= 0.658 \pm 0.044 \pm 0.047.\label{32}
\ee
Significant 
distortion of the spectrum of $e^{+}$ was observed in the KamLAND experiment.
From the  global analysis of  solar and KamLAND data for the neutrino oscillation parameters  were found 
the values \cite{Kamland}
\be
 \Delta m^{2}_{12} = 8.0^{+0.6}_{-0.4}~10^{-5}~\rm{eV}^{2};~~
\tan^{2} \theta_{12}= 0.45^{+0.09}_{-0.07}.
\label{33}
\ee

\section{Conclusion}
Strong evidence for neutrino oscillations 
was obtained in atmospheric, solar, reactor and accelerator  neutrino experiments. 
Further steps in the study of the problem of neutrino masses and mixing will include
\begin{enumerate}
\item
Detailed investigation of the discovered phenomenon.
\item
Investigation of the nature of neutrinos with definite masses (Majorana or Dirac?).
\item
Determination of the mass of the lightest neutrino.
\end{enumerate}
The value of 
the parameter $\sin^{2}\theta_{13}$ is crucial. The leptonic 
$ CP$ phase enter into the mixing matrix in the form
$U_{e3}=\sin\theta_{13}\,e^{-i\delta} $. Thus, effects of the $CP$
violation can be observed only if the parameter $\sin\theta_{13}$ is not too small.
In new reactor experiments (DOUBLE CHOOZ \cite{Chooz2}  and others) and in the accelerator 
T2K experiment \cite{T2K} significant  improvement in the  sensitivity to $\sin
^{2}\theta_{13}$ is expected. 

The establishment of the Majorana nature of $\nu_{i}$ could have
a profound importance for the understanding of the origin of small
neutrino masses.  In particular if it will be established that $\nu_{i}$ are Majorana particles it 
would be a strong indication in favor
of the famous see-saw mechanism of neutrino mass generation. 
Investigation of the neutrinoless double $\beta$-decay 
of some even-even nuclei
is the most sensitive way which could solve this problem (see reviews \cite{bbreviews}).
The probability of this process strongly 
 depends on the character of the neutrino mass spectrum, mass of the lightest neutrino and 
Majorana CP phase. Several new  ambitious experiments on the search for  neutrinoless double $\beta$-decay of different nuclei are in preparation at the moment (see \cite{bbfuture}).

I acknowledge the support of  the Italien Program  ``Rientro dei cervelli''.

\end{document}